\documentclass[aps]{revtex4}
\usepackage{amsmath}
\usepackage{amsfonts}
\usepackage{array}
\usepackage{color}

\newcolumntype{C}{>{$}c<{$}}

\begin{document}

\title{Toy models of a non-associative quantum mechanics}
\author{Vladimir Dzhunushaliev
\footnote{Senior Associate of the Abdus Salam ICTP}} 
\email{dzhun@krsu.edu.kg} \affiliation{Dept. Phys. and Microel. 
Engineer., Kyrgyz-Russian Slavic University, Bishkek, Kievskaya Str. 
44, 720021, Kyrgyz Republic}


\begin{abstract}
Toy models of a non-associative quantum mechanics are presented. The Heisenberg equation of motion is modified using a non-associative commutator. Possible physical applications of a  non-associative quantum mechanics are considered. The idea is discussed that a non-associative algebra could be the operator language for the non-perturbative quantum theory. In such approach the non-perturbative quantum theory has observables and unobservables quantities.

\end{abstract}

\maketitle

\section{Introduction}

In Ref.~\cite{jordan} the attempt was made to obtain a possible generalization of quantum mechanics on any numbers including non-associative numbers: octonions. In  Ref.~\cite{okubo1995}, the author applies non-associative algebras to physics. This book covers topics ranging from algebras of observables in quantum mechanics, to angular momentum and octonions, division algebras, triple-linear products and Yang - Baxter equations. The non-associative gauge theoretic  reformulation of Einstein's general relativity theory is also discussed. In Ref.~\cite{baez} one can find the review of mathematical definitions and physical applications for the octonions. The modern applications of the non-associativity in physics are: in Refs. \cite{Grossman}, \cite{Jackiw} it is shown that the requirement that finite translations be associative leads to Dirac's monopole quantization condition; in Ref's \cite{Gogberashvili:2005cp} and \cite{Gogberashvili:2005xb} Dirac's operator and Maxwell's equations are derived in the algebra of split-octonions. 
\par
In this paper we attempt to give toy models of a non-associative quantum mechanics using  finite dimensional non-associative algebras -- octonions or sedenions. In the previous paper \cite{Dzhunushaliev:2007cx} we have shown that in a non-associative quantum theory the observables can be presented only by elements of an associative subalgebra of a non-associative algebra of non-perturbative quantum operators. Unfortunately now we can not present any model of nonassociative quantum theory since building of such non-associative infinite dimensional algebra is very complicated mathematical problem. But in this paper we present a toy model of non-associative quantum mechanics. We can do that using an analogy with the standard quantum mechanics with the spin: if the relevant degrees of freedom for us are spin degrees of freedom only (the coordinate dependence of a wave function is not important) then we will have a qubit quantum system. The qubit quantum mechanics is much simpler the standard quantum mechanics with Pauli equation. 
\par 
In this paper we will show that there exists a finite dimensional non-associative algebra (octonions or sedenions) that has a associative subalgebra (quaternions, biquaternions). The associative subalgebra may have a non-commutative subalgebra (quaternions) and a commutative subalgebra. The quantum states of the non-commutative subalgebra are qubits. Three eigenvectors of the commutative subalgebra one can identify with three qubit fermion generations. 
\par 
Why a non-associative quantum theory can be interesting ? The reason is that it could be a candidate for a nonperturbative quantum theory formulated on the operator language. Generally speaking (according to Ref.~\cite{Dzhunushaliev:2007cx}) the elements of such algebra are unobservables but if in such non-associative algebra exists an associative subalgebra then their elements are observables. 
\par 
Thus the goal of this paper is to show that one can find a non-associative finite dimensional algebra having an associative subalgebra (which elements are observables only) and to show that one can correctly define the Heisenberg equation of motion u\emph{sing the non-associativity property}. 

\section{Non-associative quantum dynamics}
\label{naqd}

In this section we would like to present a toy model of quantum mechanics realized on a finite dimensional associative subalgebra (quanternions $\mathbb Q$ or biquaternions $\mathbb B$) of a non-associative algebra (octonions $\mathbb O$ or sedenions $\mathbb S$, respectively). 
\par 
In the usual associative quantum mechanics, we obtain the time evolution of any operator 
$\hat x$ from the Heisenberg equation of motion for the Hamiltonian $\hat H$ 
\begin{equation}
	\frac{d \hat x}{dt} = \mathrm I \left[ \hat H, \hat x \right]
\label{2a-5}
\end{equation}
where $\left[ \hat x, \hat y \right] = \hat x \hat y - \hat y \hat x$ is the commutator and $\mathrm I^2=-1$; later we will omit $\hat{ }$. Among many proposed methods for generalizing or modifying the present framework of quantum mechanics, Nambu suggested \cite{Nambu:1973qe} to modify the Heisenberg equation of motion \eqref{2a-5} into a triple product equation 
\begin{equation}
	\frac{d x}{dt} = \left\{
		h_1, h_2, x
	\right\}
\label{2a-10}
\end{equation}
where $\left\{ x,y,z \right\}$ is a triple-linear product, and we use two Hamiltonian operators $h_{1,2}$, instead of the customary one Hamiltonian as given in Eq.~\eqref{2a-5}. 
\par
Let us define a triple product following to \cite{okubo1995}. The three-linear product in a vector space $V$ can be identified by a linear mapping 
\begin{equation}
	f : V \otimes V \otimes V \rightarrow V.
\label{2a-20}
\end{equation}
For any $x,y,z \in V$, we assign an element $w \in V$, which is linear in each of $x,y,z$, and we write $w = f(x,y,z) = \{x,y,z\}$. The consistency condition 
\begin{equation}
	\frac{d }{dt} \left( x y \right) = 
	x \frac{d y}{dt} + \frac{d x}{dt} y
\label{2a-30}
\end{equation}
for Eq.~\eqref{2a-5} leads to 
\begin{equation}
	\left\{ h_1, h_2, xy \right\} = 
	x \left\{ h_1, h_2, y \right\} + 
	\left\{ h_1, h_2, x \right\} y
\label{2a-40}
\end{equation}
where $h_{1,2}, x, y \in V$ and can be any elements of the algebra $V$ (now the vector space $V$ simultaneously is an algebra). In Ref.~\cite{Dzhunushaliev:2007cx} it is shown that if we would like to introduce physical observables in a non-associative quantum theory then they can be elements of an associative subalgebra only. Therefore in contrast with the definition \eqref{2a-10} of quantum dynamic on a non-associative algebra $V$ we will propose that the  observable $x \in V_1 \subset V$  ($V_1$ is an associative subalgebra of a non-associative algebra $V$) and $h_{1,2} \in V_1 \backslash V$ are non-associative elements of the algebra $V$.

\subsection{A non-associative quantum mechanics on quaternions}
\label{qm_quaternions}

Bearing in mind that the quaternions algebra is equivalent to a qubit algebra one can apply the results of this section to the dynamics of spin, polarized photon and so on. 
\par
Let us introduce a quantum non-associative dynamics on $\mathbb{Q}$ using full non-associative algebra $\mathbb{O}$ octonions: $\mathbb{Q} \subset \mathbb{O}$ (in Appendix \ref{app1} the definitions for all algebras and multiplication table are given). For this we introduce a non-accociative commutator (n/a-commutator) in the following way 
\begin{equation}
	\left[
		i_4, i_{m+4}, b
	\right] \equiv i_4 \left( i_{m+4} b \right) - 
	\left( b i_4 \right) i_{m+4}, \quad m=1,2,3 
\label{2c-10}
\end{equation}
generalizing the usual accociative commutator $[ab, c] = (ab) c - c (ab) = abc - cab$ in the following \textcolor{red}{non-associative} way 
\textcolor{red}{
\begin{equation}
	\left[
		a, b, c
	\right] \equiv a \left( bc \right) - \left( ca \right)b , 
	\quad a,b \in \mathbb{O} , 
	\quad c \in \mathbb Q \backslash \mathbb O.
\label{2c-20}
\end{equation}
}
One can check that 
\begin{equation}
	\left[
		i_4, i_{m+4}, i_n 
	\right] = -2 \varepsilon_{mnk} i_k, \quad 
	m,n,k = 1,2,3.
\label{2c-30}
\end{equation}
The non-triviality of n/a-commutator $\left[i_4, i_{m+4}, b\right]$ is that 
\begin{equation}
	i_4 \left( i_{m+4} b \right) \neq \left( i_4 i_{m+4} \right) b 
	\quad \text{ and } 
	\left( b i_4 \right) i_{m+4} \neq b \left( i_4 i_{m+4} \right).
\label{2c-35}
\end{equation}
The consistency condition \eqref{2a-30} for Eq.~\eqref{2a-5} leads to 
\begin{equation}
	 \left[ H, x y \right] = 
	 x \left[ H, y \right] + 
	 \left[ H, x \right] y
\label{2c-50}
\end{equation}
which can be easily proved for the associative algebra. But in our case the consistency condition \eqref{2a-30} leads to 
\begin{equation}
	  \left[	i_4, i_{m+4}, bc \right] = 
	  b \left[	i_4, i_{m+4}, c \right] + 
	  \left[	i_4, i_{m+4}, b \right] c, \quad 
	  b,c = i_k \in \mathbb Q, \quad m,k=1, 2, 3
\label{2c-60}
\end{equation}
and have to be proved. Direct calculations using the Table~\ref{sedenions} (Appendix \ref{app1}) show that it is correct 
\begin{equation}
	  \left[	i_4, i_{m+4}, i_k i_l \right] = -\left[	i_4, i_{m+4}, i_l i_k \right] = 
	  i_k \left[	i_4, i_{m+4}, i_l \right] + \left[	i_4, i_{m+4}, i_k \right] i_l, 
	  \quad m,k,l = 1,2,3. 
\label{2c-75}
\end{equation}
The octonion $i_4$ can be replaced with any another octonion $n=5,6,7$. It is necessary to note that the consistency condition \eqref{2c-60} will be destroyed if the numbers 
$b,c=i_{4,5,6,7} \in \mathbb{O} \backslash \mathbb{Q}$. 
\par
For the physical application let us to introduce the following quantities 
\begin{eqnarray}
	h_{m+4} &=&  i_{m+4}\sqrt{\frac{\mathrm I \tilde\hbar}{2}}, \quad
	m = 0,1,2,3,
\label{2c-80}\\
	\hat S_k &=& i_k \frac{\mathrm I \tilde\hbar}{2}, \quad k=1,2,3
\label{2c-90}
\end{eqnarray}
here we introduce a new constant $\tilde \hbar$ as it is not evidently that in a non-associative quantum mechanics the Planck constant will be same; then the n/a-commutator will have the form 
\begin{equation}
	\left[
		h_4, h_{m+4}, \hat S_n 
	\right] = - \mathrm I \, \tilde\hbar \, \varepsilon_{mnk} \hat S_k, \quad 
	m,n,k, = 1,2,3
\label{2c-100}
\end{equation}
which should be compared with the qubit dynamic \eqref{2b-70}. Now we can define a non-associative quantum dynamic of the quantity 
$\widehat{\vec S} = s_k \hat S_k$ in an external vector field $\tilde B_m$ in the following way 
\begin{equation}
	\frac{d \widehat{\vec S}}{dt} = \mathrm I \left[
		h_4, - \left( \vec{\tilde B} \cdot \vec  \mu \right), \widehat{\vec S} 
	\right]
\label{2c-110}
\end{equation}
where $\vec{\tilde B} = \tilde B_m \vec e_m$ is an analog of a magnetic field, 
$\vec \mu = \mu h_{m+4} \vec e_m$ is an analog of a magnetic dipole for a non-associative case. Inserting $\widehat{\vec S} = s_k \hat S_k$ into \eqref{2c-110} leads to 
\begin{equation}
	\dot s_k = - \varepsilon_{mnk} \omega_m s_n
\label{2c-120}
\end{equation}
that describes the rotation of the qubit and $\omega_m = \mu B_m$ is the angular velocity. 
\par 
At the end of this section we would like to mention that full quantum mechanics on the basis of quaternions can be constructed (for details see Ref.~\cite{Adler:2001}).

\subsection{A non-associative quantum mechanics on  biquaternions}
\label{extended}

The construction similar to the subsection \ref{qm_quaternions} can be done for the biquaternions. In this case 
\begin{equation}
	\left[
		i_4, \epsilon_{m+4}, b
	\right] \equiv i_4 \left( \epsilon_{m+4} b \right) - 
	\left( b i_4 \right) \epsilon_{m+4}, \quad m=1,2,3 
\label{2d-10}
\end{equation}
One can check that 
\begin{eqnarray}
	\left[
		i_4, \epsilon_{m+4}, i_n 
	\right] &=& -2 \varepsilon_{mnk} \epsilon_k, 
\label{22-20}\\
	\left[
		i_4, \epsilon_{m+4}, \epsilon_n 
	\right] &=& 2 \varepsilon_{mnk} i_k, \quad 
	m,n,k, = 1,2,3.
\label{2d-20}
\end{eqnarray}
The non-triviality of n/a-commutator $\left[i_4, \epsilon_{m+4}, b\right]$ is that 
\begin{equation}
	i_4 \left( \epsilon_{m+4} b \right) \neq \left( i_4 \epsilon_{m+4} \right) b 
	\quad \text{ and } 
	\left( b i_4 \right) \epsilon_{m+4} \neq b \left( i_4 \epsilon_{m+4} \right).
\label{2d-30}
\end{equation}
In this case the consistency condition \eqref{2a-30} leads to 
\begin{equation}
	  \left[	i_4, \epsilon_{m+4}, bc \right] = 
	  b \left[	i_4, \epsilon_{m+4}, c \right] + 
	  \left[	i_4, \epsilon_{m+4}, b \right] c, \quad 
	  b,c = i_k, \epsilon_k \in \mathbb B, \quad m,k=1, 2, 3
\label{2d-40}
\end{equation}
and have to be proved. Direct calculations using the Table~\ref{sedenions} show that it is correct 
\begin{eqnarray}
	  \left[	i_4, \epsilon_{m+4}, i_k i_l \right] &=& 
	  -\left[	i_4, \epsilon_{m+4}, i_l i_k \right] = 
	  i_k \left[	i_4, \epsilon_{m+4}, i_l \right] + 
	  \left[	i_4, \epsilon_{m+4}, i_k \right] i_l, 
\label{2d-50}\\
	  \left[	i_4, \epsilon_{m+4}, i_k \epsilon_l \right] &=& 
	  -\left[	i_4, \epsilon_{m+4}, \epsilon_l i_k \right] = 
	  i_k \left[	i_4, \epsilon_{m+4}, \epsilon_l \right] + 
	  \left[	i_4, \epsilon_{m+4}, i_k \right] \epsilon_l, 
\label{2d-60}\\
	  \left[	i_4, \epsilon_{m+4}, \epsilon_k i_l \right] &=& 
	  -\left[	i_4, \epsilon_{m+4}, i_l \epsilon_k \right] = 
	  \epsilon_k \left[	i_4, \epsilon_{m+4}, i_l \right] + 
	  \left[	i_4, \epsilon_{m+4}, \epsilon_k \right] i_l, 
\label{2d-70}\\
	  \left[	i_4, \epsilon_{m+4}, \epsilon_k \epsilon_l \right] &=& 
	  -\left[	i_4, \epsilon_{m+4}, \epsilon_l \epsilon_k \right] = 
	  \epsilon_k \left[	i_4, \epsilon_{m+4}, \epsilon_l \right] + 
	  \left[	i_4, \epsilon_{m+4}, \epsilon_k \right] \epsilon_l, 
	  \quad m,k,l = 1,2,3. 
\label{2d-80}
\end{eqnarray}
The octonion $i_4$ can be replaced with any another octonion $i_{n+4}$ with $n=1,2,3$. 
It is necesary to note that the consistency condition \eqref{2d-40} will be destroyed if the numbers $b,c=i_{k+4}, \epsilon_{k+4}, k=1,2,3$ belong to $\mathbb{S} \backslash \mathbb{B}$. 
\par
For the physical application let us to introduce the following quantities 
\begin{eqnarray}
	h_{m+4} &=& \epsilon_{m+4}\sqrt{\frac{\mathrm I \tilde\hbar}{2}}, \quad
	m = 0,1,2,3;
\label{2d-90}\\
	\hat S_k &=& i_k \frac{\mathrm I \tilde\hbar}{2}, \quad k=1,2,3;
\label{2d-100}\\
	\hat L_k &=& \epsilon_k \frac{\mathrm I \tilde\hbar}{2}, \quad k=1,2,3
\label{2d-110}
\end{eqnarray}
then the n/a-commutator will have the form 
\begin{eqnarray}
	\left[
		h_4, h_{m+4}, \hat S_n 
	\right] &=& - \mathrm I \, \tilde\hbar \, \varepsilon_{mnk} \hat L_k, \quad 
	m,n,k, = 1,2,3;
\label{2d-120}\\
	\left[
		h_4, h_{m+4}, \hat L_n 
	\right] &=& \mathrm I \, \tilde\hbar \, \varepsilon_{mnk} \hat S_k, \quad 
	m,n,k, = 1,2,3 .
\label{2d-130}
\end{eqnarray}
Now we can define a non-associative quantum dynamic of the quantities 
$\widehat{\vec S} = s_k \hat S_k, \widehat{\vec L} = l_k \hat L_k$ in an external vector fields $\tilde B_{1,2;m}$ in the following way 
\begin{eqnarray}
	\frac{d \widehat{\vec S}}{dt} &=& \mathrm I \left[
		h_4, - n_1 \left( \vec{\tilde B}_1 \cdot \vec  \mu \right), \widehat{\vec L} 
	\right],
\label{2d-140}\\
	\frac{d \widehat{\vec L}}{dt} &=& - \mathrm I \left[
		h_4, - n_2 \left( \vec{\tilde B}_2 \cdot \vec  \mu \right), \widehat{\vec S} 
	\right]
\label{2d-150}
\end{eqnarray}
where $n_{1,2} = \pm 1$ and describe the sign of the interaction of the fields $\vec{\tilde B}_{1,2}$ with $\vec \mu_{1,2}$ and with the same definitions $\vec{\tilde B}_{1,2} = \tilde B_{1,2;m} \vec e_m$ and $\vec \mu = \mu h_{m+4} \vec e_m$ as in the previous section.
Inserting $\widehat{\vec S} = s_k \hat S_k$ and $\widehat{\vec L} = l_k \hat L_k$ into \eqref{2c-110} leads to 
\begin{eqnarray}
	\dot s_k = - \varepsilon_{mnk} \omega_m l_n,
\label{2d-160}\\
	\dot l_k = \varepsilon_{mnk} \omega_m s_n,
\label{2d-170}
\end{eqnarray}
with the same definition of $\omega_{1,2;m} = \mu \tilde B_{1,2;m}$ as the angular velocity. 

\section{Qubits quantum mechanics}

In this section we would like to present qubit system where above mentioned quantum operators from an associative subalgebra of a non-associative algebra could be operate. Here we follow to the textbook \cite{thaller}. A qubit is a quantum - mechanical two-state system. A canonical example of a qubit is provided by the spin of a spin-1/2 particle, polarized photon and so on. 
\par 
Many investigations of quantum systems do not require a ``complete'' description of the state. For example, one often neglects the position and momentum of a particle when one is only interested in the ``inner degrees of freedom'' related to the spin. This simplifies the description considerably, because the Hilbert space describing the spin of a particle with spin 1/2 is just the two-dimensional complex vector space $\mathbb{C}_2$.
\par
\textbf{Definition}: 
A quantum system with a two-dimensional Hilbert space is called a two-state system or a qubit (quantum bit). The vectors in the Hilbert space of a qubit are often called spinors.
\par 
With respect to this basis, vectors are represented by column vectors $\mathbb{C}_2$, and linear operators are represented by two-by-two matrices. For example, the basis vectors become
\begin{equation}
	\psi_+ = 
	\left( 
		\begin{array}{c}
			1 \\
			0
		\end{array}
	\right), \quad 
	\psi_- = 
	\left( 
		\begin{array}{c}
			0 \\
			1
		\end{array}
	\right)
\label{2-10}
\end{equation}
A general state of a qubit is an arbitrary superposition of the two basis states,
\begin{equation}
	\psi = c_+ \psi_+ + c_- \psi_- = 
	\left( 
		\begin{array}{c}
			c_+ \\
			c_-
		\end{array}
	\right), \text{ with }
	c_\pm \in \mathbb{C}
\label{2-20}
\end{equation}
The norm of $\psi$ and the scalar product with $\phi = d_+ \psi_+ + d_- \psi_-$ are given by
\begin{equation}
	\left\| \psi \right\|^2 = \left| c_+ \right|^2 + 
	 \left| c_- \right|^2, \quad 
	 \left\langle \psi, \phi \right\rangle = c_+^* d_+ + c_-^* d_-.
\label{2-30}
\end{equation}
Any observable has to be represented by a self-adjoint operator. With respect to a chosen orthonormal basis in the Hilbert space of a qubit, observables are thus represented by Hermitian two-by-two matrices: the three Pauli matrices 
$\vec \sigma = \left( \sigma_1, \sigma_2, \sigma_3 \right)$ which are the standard representation of the spin observables $S_1, S_2, S_3$ and as well are the representation of quaternions. 
\par 
The three Pauli matrices together with the two-dimensional unit matrix $\mathbf{1_2}$ form a basis in the four-dimensional real vector space of all Hermitian two-by-two matrices. With respect to an orthonormal basis in $\mathbb{C}^2$, any qubit observable $Q$ is represented by a linear combination of Pauli matrices 
\begin{equation}
	Q = \frac{1}{2} \left(
		a_0 \mathbf{1_2} + \sum \limits_{k=1}^3 a_k \sigma_k 
	\right)= 
	\frac{1}{2} \left(
	\begin{array}{cc}
		a_0 + a_3 	& a_1 - \mathrm I a_2\\
		a_1 + \mathrm I a_2 & a_0 - a_3
	\end{array}
	\right)
\label{2-40}
\end{equation}
with real coeffcients $a_0, \ldots, a_3$. It is necessary to note the generation relation for the algebra of Pauli matrices 
\begin{equation}
	\left[ \sigma_i , \sigma_j \right] = 
	2\mathrm I \epsilon_{ijk} \sigma_k .
\label{2-50}
\end{equation}
One can introduce the spin operators $\hat S_i = \frac{\hbar}{2} \sigma_i$ and then Eq.~\eqref{2-50} has the form 
\begin{equation}
	\left[ \hat S_i , \hat S_j \right] = 
	\mathrm I \epsilon_{ijk} \hat S_k .
\label{2-55}
\end{equation}
A general time-independent qubit Hamiltonian has the from 
\begin{equation}
	H = a_0 \mathbf{1_2} + \vec \omega \vec \sigma
\label{2b-60}
\end{equation}
where $\vec \omega = \frac{e \vec B}{mc}$ is the angular velocity speed, $e$ and $m$ is the charge and mass of a particle, $\vec B$ is a magnetic field. The dynamic of the spin 
$\widehat{\vec S} = s_i \hat S_i$ is described in the following way 
\begin{equation}
	\frac{d \widehat{\vec S}}{dt} = \mathrm I \left[ H, \widehat{\vec S} \right]
\label{2b-70}
\end{equation}
or 
\begin{equation}
	\dot s_k = \varepsilon_{ijk} \omega_i s_j
\label{2b-80}
\end{equation}
that describes the rotation of the qubit.

\section{Possible physical applications}

In the section \ref{naqd} we presented a non-associative quantum dynamic. We suppose that the  non-associative quantization procedure could be applied of a non-perturbative quantization for a field theory. The non-associative quantum dynamic presented in the section \ref{naqd} can be an approximation in this direction. Now we would like to present a few possible physical applications of such non-associative quantum dynamic. 

\subsection{Anomalous qubit rotation}

Let us consider Eq.~\eqref{2c-120} decribing the qubit rotation under action of an external constant field $\vec{\tilde B} = \left( 0,0,\tilde B \right)$ (which can be not a magnetic field in our non-perturbative case). Then we have the following equations 
\begin{eqnarray}
	\dot s_x &=& \mu \tilde B s_y,
\label{4-10}\\
	\dot s_y &=& - \mu \tilde B s_x,
\label{4-20}
\end{eqnarray}
with the solution 
\begin{eqnarray}
	s_x &=& s_{0x} \sin\left( \omega t \right),
\label{4-30}\\
	s_y &=& s_{0y} \cos\left( \omega t \right).
\label{4-40}
\end{eqnarray}
The solution describes the rotation of qubit around the external constant field $\vec{\tilde B}$ in the plane $xy$. Comparing Eq's~\eqref{2c-120} and \eqref{2b-80} we see that the rotation of qubit in the non-associative quantum mechanics is in the opposite direction in comparison with the rotation of spin around the magnetic field in the standard quantum mechanics. 
\par 
For the extended version of qubit presented in subsection \ref{extended} we use Eq's~\eqref{2d-160}~\eqref{2d-170} with 
\begin{eqnarray}
	\tilde B_{1,z} &=& \tilde B_1, \quad \omega_{1,z} = \omega_1,
\label{4-50}\\
	\tilde B_{2,z} &=& \tilde B_2, \quad \omega_{2,z} = \omega_2,
\label{4-60}\\
	\dot s_x &=& \omega_1 l_y,
\label{4-70}\\
	\dot l_y &=& \omega_2 s_x
\label{4-80}
\end{eqnarray}
with following solution
\begin{equation}
	s_x = s_{0x} e^{\pm t \sqrt{n_1 n_2 \omega_1 \omega_2}}, 
	l_y = l_{0y} e^{\pm t \sqrt{n_1 n_2 \omega_1 \omega_2}}.
\label{4-90}
\end{equation}
It seems that the solution with $n_1 n_2 > 0$ is physically senseless as in this case we have exponentially increasing/decreasing operators $\hat L, \hat S$. But in the case $n_1 n_2 < 0$ we have the rotation of extended qubits in a plane.

\subsection{Fermion qubit generations}

Let us consider the commutative and associative subagebra $\mathbb A$ spanned on basis $\left( 1,i_3,\epsilon_3,i_0 \right)$ which is the commutative subalgebra of the noncommutative and associative algebra of biquaternions $\mathbb A \subset \mathbb{B} \subset \mathbb S$ (for the defenitions of biquaternions and sedenions see Appendix \ref{app1}). The matrix representation of $\mathbb{A}$ is 
\begin{alignat}{4}
	i_3 = &\mathrm{I} \left(
	\begin{array}{cc}
		-\sigma_3 	& 0 \\
		0						& 1
	\end{array}
	\right) & = 
	\mathrm{I} & \left(
	\begin{array}{ccc}
		-1 	& 0 	&	0	\\
		0		&	1		& 0	\\
		0		&	0		&1
	\end{array}
	\right)
\label{2b-2}\\
	\epsilon_3 = &\left(
	\begin{array}{cc}
		\sigma_3 		& 0 \\
		0						& 1
	\end{array}
	\right) & = & 
	\left(
	\begin{array}{ccc}
		1 	& 0 	&	0	\\
		0		&	-1		& 0	\\
		0		&	0		&1
	\end{array}
	\right)
\label{2b-4}\\
	i_0 = &\mathrm{I} \left(
	\begin{array}{cc}
		-\mathbf{1_2} 	& 0 \\
		0								& 1
	\end{array}
	\right) & = 
	\mathrm I & \left(
	\begin{array}{ccc}
		-1 	& 0 	&	0	\\
		0		&	-1		& 0	\\
		0		&	0		&1
	\end{array}
	\right).
\label{2b-10}
\end{alignat}
The basis vectors are
\begin{equation}
	\xi_1 = 
	\left( 
		\begin{array}{c}
			1 \\
			0	\\
			0
		\end{array}
	\right), \quad 
	\xi_2 = 
	\left( 
		\begin{array}{c}
			0 \\
			1	\\
			0
		\end{array}
	\right), \quad 
	\xi_3 = 
	\left( 
		\begin{array}{c}
			0 \\
			0	\\
			1
		\end{array}
	\right)
\label{2b-40}
\end{equation}
which are eigenvectors of matrixes \eqref{2b-2}-\eqref{2b-10}. For such system we can apply Heisenberg equation of motion \eqref{2d-140} \eqref{2d-150}. 
\par
One can say that the index $k$ (by $\xi_k$) enumerate ``a fermion generation'' of extended qubits living in the vector space $E$ of the matrix representation of biquaternions. The vector space $E_1$ spanned on the basis vectors \eqref{2b-40} is a vector subspace $E_1 \subset E$. According to equation of motion \eqref{2d-160} \eqref{2d-170} the generations of extended qubits can mix up. But it is correct only if there exists the interaction term 
$\left( - \vec{\tilde B}_2 \cdot \vec  \mu \right)$ in the opposite case the qubit generations can not be mix up. 
\par 
In the standard model of particle physics \cite{SM} there are open questions 
which have not yet found an answer. Chief among these is the fermion family or generation 
puzzle as to why the first generation of quarks and leptons (up quark, down quark, 
electron and electron neutrino) are replicated in two families or generations of increasing  mass (the second generation consisting of charm quark, strange quark, muon and muon 
neutrino; the third generation consisting of top quark , bottom quark, tau and tau neutrino).
One can presuppose that a non-associative infinite dimensional quantum theory may shed light on the generation puzzle of fermions. 

\subsection{Slave-boson decomposition}

The most important for the toy model of a non-associative quantum mechanics presented here is the factorization of Hamiltonian in the sense that instead of usual commutator
\begin{equation}
	\left[ H,x \right] = Hx - xH
\label{4c-10}
\end{equation}
we use a non-associative commutator 
\begin{equation}
	\left[
		h_4, h_{m+4}, x
	\right] = h_4 \left( h_{m+4} x \right) - 
	\left( x h_4 \right) h_{m+4} .
\label{4c-20}
\end{equation}
Roughly speaking one can say that the Hamiltonian $H$ is factorized on two non-associative factors  $h_4, h_{m+4}$. One can try to find some connection of such decomposition with something similar in physics. In this connection one can think of ``slave-boson decomposition''  in the $t-J$ model of High-T$_c$ superconductivity.
\par 
It is widely believed that the low energy physics of High-T$_c$ cuprates is described in terms of $t$-$J$ type model, which is given by \cite{LN9221}
\begin{equation}
	H = \sum \limits_{i,j} J\left(
	{{S}}_{i}\cdot {{S}}_{j}-\frac{1}{4} n_{i} n_{j} \right)
	-\sum_{i,j} t_{ij}
	\left(c_{i\sigma}^\dagger
	c_{j\sigma}+{\rm H.c.}\right)
\label{4c-30}
\end{equation}
where $t_{ij}=t$, $t'$, $t''$ for the nearest, second nearest and 3rd nearest
neighbor pairs, respectively. The effect of the strong Coulomb repulsion is
represented by the fact that the electron operators $c^\dagger_{i\sigma}$ and
$c_{i\sigma}$ are the projected ones, where the double occupation is
forbidden.  This is written as the inequality
\begin{equation}
	\sum_{\sigma} c^\dagger_{i\sigma} c_{i \sigma} \le 1 
\label{4c-40}
\end{equation}
which is very difficult to handle.  A powerful method to treat this constraint
is so called the slave-boson method \cite{B7675,C8435}. In this approach the electron operator is represented as
\begin{equation}
	c^\dagger_{i\sigma} = f_{i\sigma}^\dagger b_{i} 
\label{4c-50}
\end{equation}
where $f_{i\sigma}^\dagger$, $f_{i \sigma}$ are the fermion operators, while $b_{i}$ is the
slave-boson operator. This representation together with the constraint 
\begin{equation}
	f_{i\uparrow}^\dagger f_{i\uparrow} + f_{i\downarrow}^\dagger f_{i\downarrow} + 
	b^\dagger_{i} b_{i} = 1
\label{4c-60}
\end{equation}
reproduces all the algebra of the electron operators. The physical meaning of the operators $f$ and $b$ is unclear: do exist these fields or not ? 
\par 
If we compare the factorizations \eqref{4c-50} and \eqref{4c-20} one presuppose that the operators $f_{i\sigma}^\dagger, b_{i}$ are elements of an infinite dimensional non-associative algebra $\mathfrak Q$. This algebra has an associative subalgebra 
$\mathfrak A \subset \mathfrak Q$ and the operator $c^\dagger_{i\sigma} \in \mathfrak A$ is observable but the operators $f_{i\sigma}^\dagger, b_{i} \in \mathbb Q \backslash \mathfrak A$ are unobservables. It could mean that the High-T$_c$ superconductivity (similar to quantum chromodynamics) can be understood on the basis of a \emph{non-perturbative} quantum theory and one can assume that the  non-perturbative quantum theory (on the operator language) could be realized as a non-associative quantum theory (realized as a non-associative algebra 
$\mathfrak Q$) with observables belonging to an associative subalgebra $\mathfrak A$ and unobservables belonging to $\mathfrak Q \backslash \mathfrak A$. 
\par 
The difference between the slave-boson decomposition \eqref{4c-50} and the non-associative commutator \eqref{4c-20} is that the first one is a quantum theory where the coordinate degrees of freedom are taking into account, but the second one is a quantum mechanics where the coordinate degrees of freedom are not taking into account. 
\par 
It is necessary to note here that in Ref's~\cite{Niemi:2005qs}-\cite{oma1} there are a classical generalization of slave-boson decomposition on gauge theories, so called -- 
``spin-charge separation''. 

\section{Outlook}

Thus we have shown that one can generalize the standard finite dimensional quantum mechanics (for example, qubit quantum mechanics) to a non-associative finite dimensional quantum mechanics realized on a finite dimensional associative algebra which is a subalgebra of a non-associative algebra. We have considered two cases: quaternionic and biquaternionic non-associative qubit quantum mechanics. In both cases the non-associativity is realized in the Heisenberg quantum equation of motion: on RHS of corresponding equations the usual commutator is changed on a non-associative commutator. 
\par 
The biquaternion version of non-associative quantum mechanics has commutative 
$(1, i_0, \epsilon_3, i_3)$ and noncommutative $(1,i_1,i_2,i_3)$ observables. It allow us to suppose that an infinite dimensional non-associative quantum theory will have an associative subalgebra having physical observables with commutators 
$(\hat a \hat a^\dagger - \hat a^\dagger \hat a = \mathrm I \hbar)$ and anticommutators 
$(\hat f \hat f^\dagger + \hat f^\dagger \hat f = \mathrm I \hbar)$. Probably it means that the unification of bosons and fermions can be done not only on the basis of supersymmetry but in a non-associative quantum theory as well. 
\par 
Now we would like to list the results and properties of discussed here a non-associative quantum mechanics:
\begin{itemize}
	\item Two examples of a finite dimensional quantum mechanics are presented.
	\item The Heisenberg quantum equation of motion are essentially non-associtive.
	\item The non-associativity leads to the fact that the usual Hamiltonian can not be introduced as the product of two operators.
	\item Generally speaking the non-associative factors $(i_4, i_{m+4})$ or $(i_4, \epsilon_{m+4})$ are unobservable physical quantities that remind hidden parameters in the theory with hidden parameters.
	\item The non-associative quantum theory can be alternative one to supersymmetric theories.
	\item In Ref.~\cite{Dzhunushaliev:2007cx} it is shown that the non-associative quantum theory can describe non-local objects like strings and so on.
\end{itemize}

\appendix
\section{Sedenions}
\label{app1}

Sedenions \cite{Carmody} form an algebra with non-associative but alternative multiplication and a multiplicative modulus. It consists of one real axis (to basis $1$), eight imaginary axes (to bases $i_n$ with $i_n^2 = -1, n=0, \ldots , 7$), and seven real axes (to bases $\epsilon_n$ with $\epsilon_n^2 = +1, n=1, \ldots , 7$). The multiplication table is given in Table~\ref{sedenions}. The sedenions non-associative algebra contains following important subalgebras: 
\begin{itemize}
	\item the associative quaternion subalgebra $\mathbb Q$ with $i_n, n = 1,2,3$;
	\item the associative biquaternion subalgebra $\mathbb B$ with 
	$i_0, i_n, \epsilon_n, n=1,2,3$;
	\item the non-associative octonion subalgebra $\mathbb O$ with $i_n, n= 0, \ldots 7$.
\end{itemize}
\begin{table}[h]
\begin{tabular}{|
C|| C|| C| C| C| C| C| C| C|| C|| C| C| C| C| C| C| C| C|}                                                
\hline
			&	  1  					&  i_1  				&  i_2  				&  i_3   				&		  i_4  			&  i_5  				&  i_6   				&	 i_7  				&	 i_0   				
			&  \epsilon_1  	&  \epsilon_2  	&  \epsilon_3   &  \epsilon_4  	&  \epsilon_5  	& 							 \epsilon_6   	&	 \epsilon_7  	  
\\ 
\hline \hline
1			&		  1  				&  i_1  					&  i_2  				&  i_3   				&		  i_4  			&  i_5  			&  i_6   				&	 i_7  					&	 i_0   				
			&  \epsilon_1  	&  \epsilon_2  		&  \epsilon_3   &  \epsilon_4  	&  \epsilon_5  	& 							 \epsilon_6   	&	 \epsilon_7  	  
\\ 
\hline \hline
i_1		&	 i_1  				&  -1  						&  i_3  				&  -i_2   			&		  i_5  			& -i_4 				&  -i_7   			&	 i_6 	&	 -\epsilon_1   					&  i_0  				&  \epsilon_3  
			&  -\epsilon_2  &  \epsilon_5  		&  -\epsilon_4  &	 -\epsilon_7  &	 \epsilon_6   
\\ 
\hline 
i_2		&	 i_2  				&  -i_3  					&  -1  					&  i_1   				&		  i_6  			
			& i_7 					&  -i_4   				&	 -i_5 				&	 -\epsilon_2  &  -\epsilon_3  							&  i_0  				&  \epsilon_1 		&  \epsilon_6  	&  \epsilon_7  	&	 -\epsilon_4  
			&	 -\epsilon_5   
\\ 
\hline 
i_3 	&	 i_3  				&  i_2  					&  -i_1 				&  -1  					&  i_7   				
			&	-i_6 & i_5 		&  -i_4   				&	 -\epsilon_3 	&	 \epsilon_2  	&  -\epsilon_1  
			&  i_0  				&  \epsilon_7 		&  -\epsilon_6  &  \epsilon_5  	&	 -\epsilon_4    
\\ 
\hline 
 i_4 	&	 i_4  				&  -i_5  					&  -i_6 				&  -i_7  					
			&  -1   				&	 i_1 						& i_2 					&  i_3   		
			&	 -\epsilon_4 	 
			&	 -\epsilon_5  &  -\epsilon_6  	&  -\epsilon_7 	&  i_0  
			&  \epsilon_1 	&  \epsilon_2  		&  \epsilon_3  		
\\ 
\hline 
 i_5 	&	 i_5  				&  i_4  					&  -i_7 				&  i_6  					
			&  -i_1   			&	 -1 						& -i_3 					&  i_2   		
			&	 -\epsilon_5 	 
			&	 \epsilon_4  	&  -\epsilon_7  	&  \epsilon_6 	&  -\epsilon_1  
			&  i_0 					&  -\epsilon_3  	&  \epsilon_2  		
\\ 
\hline 
 i_6 	&	 i_6  				&  i_7  					&  i_4 				&  -i_5  					
			&  -i_2   			&	 i_3 						& -1 					&  -i_1   		
			&	 -\epsilon_6 	 
			&	 \epsilon_7  	&  \epsilon_4  		&  -\epsilon_5 &  -\epsilon_2  
			&  \epsilon_3 	&  i_0  					&  -\epsilon_1  		
\\ 
\hline 
 i_7 	&	 i_7  				&  -i_6  					&  i_5 				&  i_4  					
			&  -i_3   			&	 -i_2 					& i_1 				&  -1   		
			&	 -\epsilon_7 	 
			&	 -\epsilon_6  &  \epsilon_5  		&  \epsilon_4 &  -\epsilon_3  
			&  -\epsilon_2 	&  \epsilon_1  		&  i_0  		
\\ 
\hline \hline
 i_0 	&	 i_0  				&  -\epsilon_1  	&  -\epsilon_2 &  -\epsilon_3  					
			&  -\epsilon_4  &	 -\epsilon_5 		& -\epsilon_6  &  -\epsilon_7   		
			&	 -1 	 
			&	 i_1  				&  i_2  					&  i_3 					&  i_4  
			&  i_5 					&  i_6  					&  i_7  		
\\ 
\hline \hline
 \epsilon_1 	
			&	 \epsilon_1  	&  i_0  					&  \epsilon_3 	&  -\epsilon_2  					
			&  \epsilon_5  	&	 -\epsilon_4 		& -\epsilon_7 	&  \epsilon_6   		
			&	 i_1 	 
			&	 1  					&  -i_3  					&  i_2 					&  -i_5  
			&  i_4 					&  i_7  					&  -i_6  		
\\ 
\hline 
 \epsilon_2 	
			&	 \epsilon_2  	&  -\epsilon_3  	&  i_0 					&  \epsilon_1  		
			&  \epsilon_6  	&	 \epsilon_7 		& -\epsilon_4 	&  -\epsilon_5   		
			&	 i_2 	 
			&	 i_3  				&  1  						&  -i_1 				&  -i_6  
			&  -i_7 				&  i_4  					&  i_5  		
\\ 
\hline 
 \epsilon_3 	
			&	 \epsilon_3  	&  \epsilon_2  		&  -\epsilon_1 	&  i_0  		
			&  \epsilon_7  	&	 -\epsilon_6 		& \epsilon_5 		&  -\epsilon_4   		
			&	 i_3 	 
			&	 -i_2  				&  i_1  					&  1 						&  -i_7  
			&  i_6 					&  -i_5  					&  i_4  		
\\ 
\hline 
 \epsilon_4 	
			&	 \epsilon_4  	&  -\epsilon_5  	&  -\epsilon_6 	&  -\epsilon_7  		
			&  i_0  				&	 \epsilon_1 		& \epsilon_2 		&  \epsilon_3   		
			&	 i_4 	 
			&	 i_5  				&  i_6  					&  i_7 					&  1  
			&  -i_1 				&  -i_2  					&  -i_3  		
\\ 
\hline 
 \epsilon_5 	
			&	 \epsilon_5  	&  \epsilon_4  		&  -\epsilon_7 	&  \epsilon_6  		
			&  -\epsilon_1  &	 i_0 						& -\epsilon_3 	&  \epsilon_2   		
			&	 i_5 	 
			&	 -i_4  				&  i_7  					&  -i_6 				&  i_1  
			&  1 						&  i_3  					&  -i_2  		
\\ 
\hline 
 \epsilon_6 	
			&	 \epsilon_6  	&  \epsilon_7  		&  \epsilon_4 	&  -\epsilon_5  		
			&  -\epsilon_2  &	 \epsilon_3 		& i_0 					&  -\epsilon_1   		
			&	 i_6 	 
			&	 -i_7  				&  -i_4  					&  i_5 					&  i_2  
			&  -i_3 				&  1  						&  i_1  		
\\ 
\hline 
 \epsilon_7 	
			&	 \epsilon_7  	&  -\epsilon_6  	&  \epsilon_5 	&  \epsilon_4  		
			&  -\epsilon_3  &	 -\epsilon_2 		& \epsilon_1 		&  i_0   		
			&	 i_7 	 
			&	 i_6  				&  -i_5  					&  -i_4 				&  i_3  
			&  i_2 					&  -i_1  					&  1  		
\\ 
\hline 
\end{tabular}
\caption{The sedenions multiplication table.} 
\label{sedenions}
\end{table}

\end{document}